# Optimizing compositional and atomic-level information of oxides in atom probe tomography

K. A. Hunnestad[1], C. Hatzoglou[1], F. Vurpillot[5], I-E. Nylund[1], Z. Yan[2,3], E. Bourret[3], A. T. J. van Helvoort[4], D. Meier[1*]

[1]Department of Materials Science and Engineering, Norwegian University of Science and Technology (NTNU), 7491 Trondheim, Norway

[2]Department of Physics, ETH Zurich, Zürich, Switzerland

[3]Materials Sciences Division, Lawrence Berkeley National Laboratory, Berkeley, CA, USA

[4]Department of Physics, Norwegian University of Science and Technology (NTNU), 7491 Trondheim, Norway

[5]UNIROUEN, CNRS, Groupe de Physique des Matériaux, Normandie Université, Av. de l'Université BP12, 76801 St Etienne du Rouvray, France

*Corresponding author. Email: dennis.meier@ntnu.no

## Abstract

**Atom probe tomography (APT) is a 3D analysis technique that offers unique chemical accuracy and sensitivity with sub-nanometer spatial resolution. Recently, there is an increasing interest in the application of APT to complex oxides materials, giving new insight into the relation between local variations in chemical composition and emergent physical properties. However, in contrast to the field of metallurgy, where APT is routinely applied to study materials at the atomic level, complex oxides and their specific field evaporation mechanisms are much less explored. Here, we perform APT measurements on the hexagonal manganite $ErMnO_3$ and systematically study the effect of different experimental parameters on the measured composition and atomic structure. We demonstrate that both the mass resolving power (MRP) and compositional accuracy can be improved by increasing the charge-state ratio (CSR) working at low laser energy (< 5 pJ). Furthermore, we observe a substantial preferential retention of Er atoms, which is suppressed at higher CSRs. We explain our findings based on fundamental field evaporation concepts, expanding the knowledge about the impact of key experimental parameters and the field evaporation process in complex oxides in general.**

# 1. Introduction

Laser-assisted atom probe tomography (APT) is a well-established characterization technique in metallurgy, where it enabled numerous breakthroughs, shaping, e.g., the modern understanding of clustering and chemical segregation[1,2]. In addition, APT was applied to a range of material systems and geometries, ranging from frozen water[3] to human enamel[4], showing its broad application potential. Recently, there is an increasing interest in measurements on oxide materials, as APT can provide unprecedented insight into key characteristics, such as compositional variations, dopant distributions and defects, that underpin their functional properties. For example, APT has been applied to quantify impurities at grain boundaries in $CeO_2$[5], to analyze site-specific doping in oxide semiconductors [6], and to investigate the correlation between polar discontinuities and chemical composition at grain boundaries in ferroelectric polycrystals[7]. Furthermore, APT has been used to access chemical disorder at the atomic level [6,8–11], holding great potential for future studies of emergent phenomena in complex oxides.

Despite the remarkable opportunities, APT measurements also impose different challenges and a range of evaporation artefacts can arise, leading to systematic biases, such as preferential evaporation, retention, co-evaporation, and molecular dissociation[12–17]. These effects can obscure the data analysis, resulting in, e.g., inaccurate compositional measurements, loss of information, or spatial aberrations. To achieve accurate chemical measurements, it is crucial to optimize the experimental parameters; this includes the base temperature, detection rate, laser energy and frequency, as well as the shape of the sample under investigation, which controls the electric field strength at the surface and, hence, the evaporation mechanisms. The importance of the experimental parameters is reflected by several studies, which investigated their influence on the data quality and chemical bias[12,13,18–29]. The laser energy determines the change in temperature during pulsing and is varied to meet the field evaporation criteria for all elements. For many materials, including metals, a laser energy in the range of 30-100 pJ has been demonstrated to be adequate. The base temperature is usually set as low as possible to minimize thermal effects, and high laser frequencies and detection rates are favorable in order to reduce noise and unwanted evaporation events between pulses. As higher temperatures reduce the electric field required for field evaporation, the charge-state ratio (CSR) of a selected ionic species is often used as an indicator for the electric field strength at the apex. The CSR comprises different analysis parameters that are crucial for the field evaporation (e.g., tip shape, base temperature, laser pulse energy, and detection rate) and, hence, represents a important metric for understanding field evaporation processes[30]. Along with the other experimental parameters, the CSR is typically optimized so that the APT analysis yields an accurate composition measurement for a selected reference system, as well as sharp peaks in the mass spectrum to achieve optimal mass resolving power (MRP).

Interestingly, several oxides have been reported to exhibit a completely different trend with respect to the common analysis parameters and the best results were achieved with low instead of high laser energies (sometimes as low as 0.2 pJ)[31–34]. One reason for this is the low thermal conductivity generally found in oxides, which leads to so-called thermal tails in the mass spectrum. The latter can be minimized by decreasing the laser pulse energy. Furthermore, neutral $O_2$ molecules can be desorbed on the surface preventing ionization, which usually goes undetected and leads to a loss of information and apparent oxygen deficiency[35]. Here, a lower laser energy is beneficial as it reduces the temperature and increases the ionization probability of neutral $O_2$ molecules. So far, however, the effect of very low pulse energies in APT measurements has been investigated only in a limited number of oxide materials. In addition, comprehensive investigations concerning the impact on the spatial resolution and atomic imaging in oxide materials are scarce.

Here, we perform a systematic analysis using hexagonal erbium manganite (ErMnO$_3$), monitoring how different APT experimental parameters affect the measurement. ErMnO$_3$ has been intensively studied for its multiferroicity[36–38] and unusual ferroelectric domain wall properties[39,40]. Its atomic structure is well known[41–43], making it an ideal model system. The unit cell consists of alternating Mn and Er lattice planes with a spacing of 5.71 Å along the [001] zone axis, which can readily be resolved by APT[6]. We investigate the effect of the laser energy on the MRP and the measured composition, showing that lower laser energies yield better results, consistent with previous observations on other oxide systems. Going beyond compositional analysis on the bulk level, we demonstrate that variations in CSR (or laser energy) cause substantial variations in the measured distances between the atomic planes of Er and Mn. This CSR-dependent atomic offset is explained based on the preferential retention of Er atoms and changes in specimen temperature, giving new insight into the field evaporation process in complex oxides.

## 2. Material and Methods

The single crystals of ErMnO$_3$ (space group P6$_3$cm) used in this study are grown by the pressurized floating-zone method[44]. Samples are first cut from a larger crystal and then oriented by Laue diffraction to gain surfaces perpendicular to the hexagonal c-axis of the material. A Helios NanoLab DualBeam focused ion beam (FIB) with a Ga source is used to prepare APT needle specimens from the bulk samples as described in Ref. [45].

High-resolution scanning transmission electron microscopy (STEM) data is taken using a JEOL JEM ARM200F operating at 200kV. Sample checks and diffraction measurements are taken with a JEOL JEM-2100F Field Emission Electron Microscope, also operated at 200kV. All TEM data is obtained directly from the APT specimens with no further sample preparation before the APT analysis.

APT analysis is done with a Cameca LEAP 5000XS, operated in laser pulsing mode with a frequency of 250 kHz and detection rate at 0.5% (i.e., 1 atom detected on average every 200 laser pulse). The base temperature is either 25K or 50 K as detailed in the figure captions and the laser energy is set to 5 pJ unless specified differently. To reconstruct the raw APT data into 3D datasets, the software Cameca IVAS 3.6.12 is used. Radius evolution is determined based on the voltage profile and correct reconstruction parameters are found by measuring the plane-to-plane distances along the 001 pole. Typically, an image compression factor of 1.8 is used and a field reduction factor of 2.7 with 30 V/nm as the evaporation field (as for bulk Mn). MRP values are obtained from IVAS by calculating $M/\Delta M$, where $\Delta M$ is the full width at half maximum (FWHM) of the largest peak in the mass spectrum. For the SDM analysis we use the Norwegian Atom Probe App (NAPA) software[46], developed in MATLAB® which calculates the SDM along the direction normal to the atomic planes for best signal-to-noise ratio.

## 3. Results and Discussion
### 3.1 Sample preparation

A representative FIB-cut ErMnO$_3$ specimen as used for the APT analysis is shown in Figure 1(a). TEM is applied to verify structural integrity after FIB preparation. Figure 1(b) presents a high-angle annular dark-field scanning TEM (HAADF-STEM) lattice image gained at the tip of the needle, showing the alternating atomic planes of Er and Mn with the characteristic up-up-down displacement of the Er atoms[47–49] in the [100] direction. Only a thin amorphous layer with a thickness of about 10 nm is observed (not shown) on the outside of the needle specimen. In Figure 1(c), a selected area electron diffraction (SAED) pattern is displayed, corresponding to the whole tip volume; only Bragg reflections associated with the [100] zone axis are observed, excluding any FIB-induced secondary phases. The characteristic needle shape of the specimens is necessary to focus the electric field at the tip and achieve the field strength required for evaporation (around 30 V/nm). Following the evaporation of the region of interest, the final 3D chemical map in Figure 1(d) is then obtained after reconstruction.

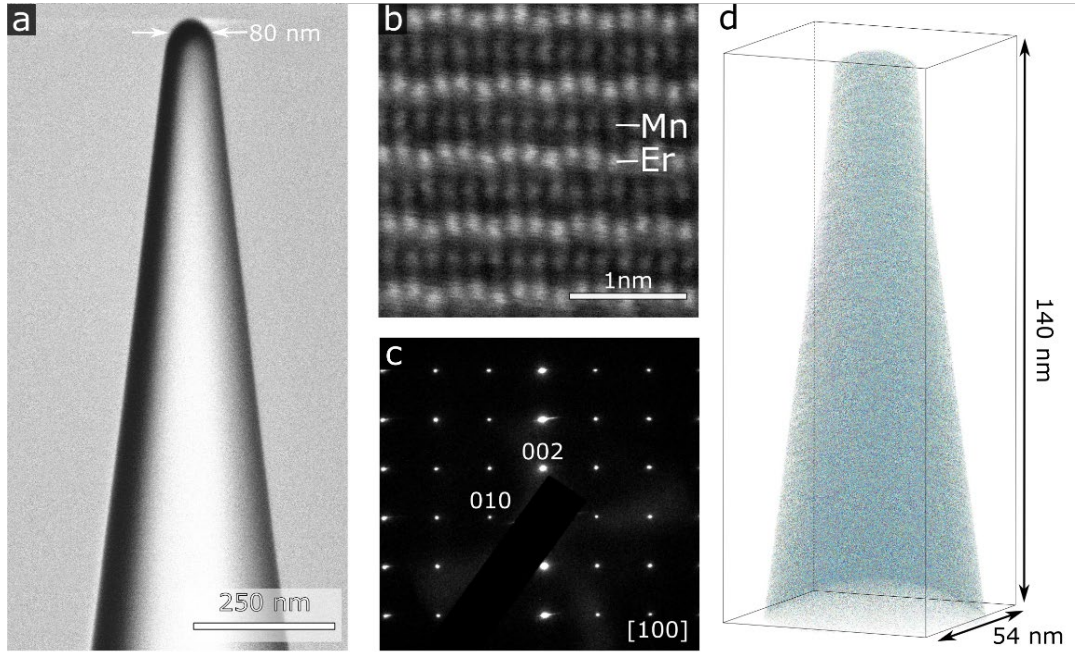

**Figure 1: Sample preparation and atom probe tomography on ErMnO$_3$.** (a) Scanning electron microscopy (SEM) image of an ErMnO$_3$ APT specimen, shaped into a nanoscale needle using a FIB. (b) HAADF-STEM lattice image showing alternating Er and Mn planes and (c) SAED pattern obtained from the tip of an APT needle, indicating crystallinity. Data is taken along the [100] zone axis. (d) 3D virtual specimen obtained after reconstructing the raw data from the field evaporation. O atoms are colored blue, Mn yellow and Er red.

### 3.2 Parameter optimization

For achieving optimal experimental conditions in APT, the laser energy is critical. Its influence on the MRP can be seen in Figure 2(a), where parts of the mass spectrum are shown gained with two laser energies (5 pJ and 30 pJ). The peaks are narrower for the lower laser energy, indicating a higher MRP and, hence, more accurate chemical quantification and sensitivity. In Figure 2(b), it is shown in more detail that lower laser energies ($\approx$ 1 pJ) indeed result in higher MRP. Furthermore, it is shown in Figure 2(c) that the CSR of the ErO species $\left(\frac{ErO^{2+}}{ErO^{2+}+ErO^{+}}\right)$ ranges from 0.2 to 0.9 when increasing the laser energy from 0.2 pJ to 200 pJ[50]. This change in CSR is also reflected in the measured composition as seen in Figure 2(d). For higher CSR, the composition approaches the nominal values of 20% Er, 20% Mn, and 60% O. Note that the data in each of the plots in Figure 2(b)–(d) are obtained from three different specimens to exclude bias from a specific sample geometry.

The trend we observe in our APT measurements on ErMnO$_3$ is consistent with previous reports on other oxide materials, that is, for high CSR and, thus, high electric field strength at the surface, the measured oxygen concentration is higher[51]. This is because fewer oxygen leaves as neutral O$_2$ molecules, which are undetectable in the experiment. At the same time, the Mn concentration drops

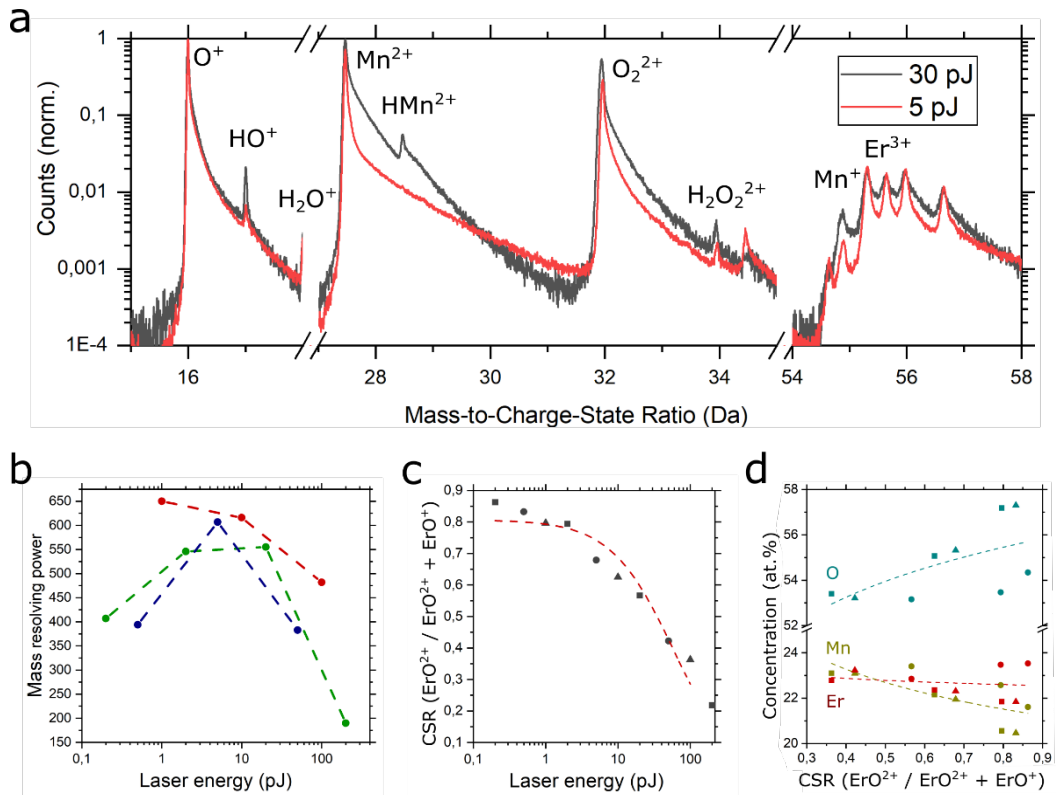

**Figure 2: Influence of APT parameter on key analysis metrics.** (a) Mass spectrum of ErMnO$_3$ for two different laser energies, 5pJ and 30pJ. Thermal tails are reduced at 5pJ compared to 30pJ. (b) MRP as a function of the laser energy from three different APT specimens (distinguished by colors). All data is acquired under the same parameters aside from the laser energy (i.e. base temperature: 50K, detection rate: 0.5%, pulse frequency: 250 kHz). (c) CSR evolution of the ErO ionic species (ErO$^{2+}$ and ErO$^+$) as a function of the laser energy. (d) Composition of the full data volume as a function of CSR. Increasing CSR corresponds to reducing the laser energy. All three plots (b-d) show that analysis conditions are improved towards lower laser energies (<10pJ).

below the Er concentration, possibly due to preferential evaporation of Mn.

### 3.3 Crystallographic poles and atomic resolution

Some materials form facets perpendicular to a zone axis on the surface of the APT needle during evaporation. These regions, called crystallographic poles, facilitate excellent spatial resolution and can be used to visualize the atomic planes along the zone axis of the pole. In the case of oriented ErMnO$_3$, a pole is formed along the [001] zone axis as displayed in Figure 3(a), which corresponds to a 2D slice of the APT dataset. Figure 3(b) presents the atomic planes of Mn. In Figure 3(c), a profile across the pole of the composition is shown; the data shows that there is a clear chemical bias in the pole region as expected from the different field conditions at the pole[52,53].

In the pole region, the atomic planes of the different elements can be resolved as summarized in Figure 4(a)-(c). While the O planes are not directly visible, the alternating planes of Mn and Er are readily resolved, consistent with HAADF-STEM data in Figure 1(b) (Figure 4(d) shows a schematic illustration). For a more quantitative analysis, spatial distribution maps (SDMs) are calculated for all the elements using the Mn ions as reference (Figure 4(e))[54]. Because ions of the same type have overlapping atomic planes, the Mn-Mn SDM can be used to probe the plane-to-plane distance and calibrate the reconstruction so that the distance becomes equal to $d_{Mn}$ = 5.7 Å. The Mn-O SDM is not shown as it reflects the evaporation conditions rather than the actual positions; O close to the Er planes tend to evaporate as ErO species, whereas O from the Mn planes evaporate as O. Finally, when considering the Mn-Er SDM, we find that the peaks are located between the Mn planes as we expect from the unit cell structure. When analyzing the exact position more carefully, however, we find a deviation of about 0.04 $d_{Mn}$ from the established position of the Er layer [55].

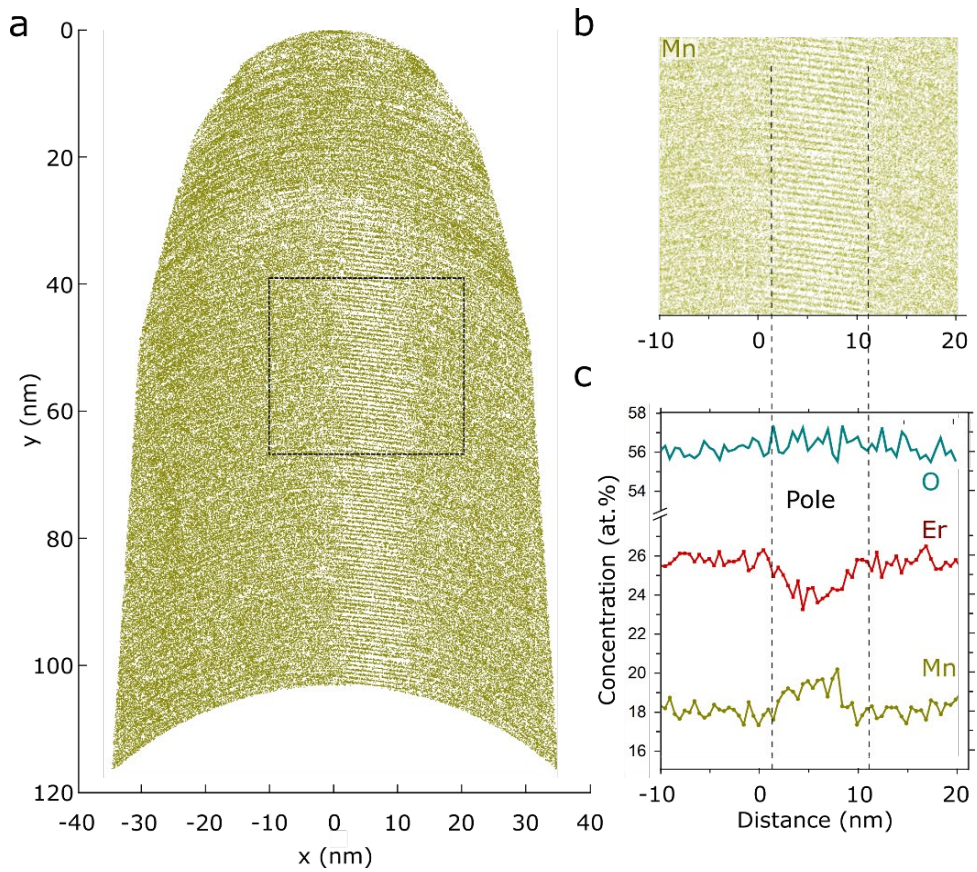

**Figure 3: Influence of crystallographic pole on the chemical composition**. (a) Slice of the full dataset containing the pole, where only the Mn ions are displayed. (b) Smaller region of interest containing the pole, corresponding to the position marked in (a). (c) The chemical composition measured from left to right in the region seen in (b) are shown. A chemical bias around the pole can be seen through an increase in Mn concentration and a decrease in Er concentration. Data is taken at 25K.

### 3.4 Atomic planes offset

To understand the origin of the Mn-Er offset observed in the SDMs, we analyze the effect of varying electric field strength represented by the CSR. The change in electric field strength is achieved by varying the laser pulse energy as shown in Figure 5. The data shows a clear trend, approaching the expected Mn-Er offset of 0.5 $d_{Mn}$ with increasing CSR. In contrast, as the CSR decreases (increasing laser energy and temperature), the offset is reduced and the measured location of the Er planes approaches that of the Mn planes.

An offset with respect to the expected position of atomic planes was observed previously for one-phase binary alloys as discussed in Ref. [55]. Samples can change their surface roughness locally (or curvature) so that the local electric field strength remains constant. This effect depends on the evaporation rate of the different species and will eventually lead to a distinct and constant tip shape. The roughness (or curvature) above two different surface atoms will then be proportional to the species evaporation fields. Thus, Er and Mn atomic planes will exhibit different local roughness, resulting in trajectory aberrations and spatial deviations relative to each other (offset). Within this model, the observed deviation (Fig. 5) relates to an increase in difference between the effective evaporation fields of Er and Mn atomic planes as the laser energy increases.

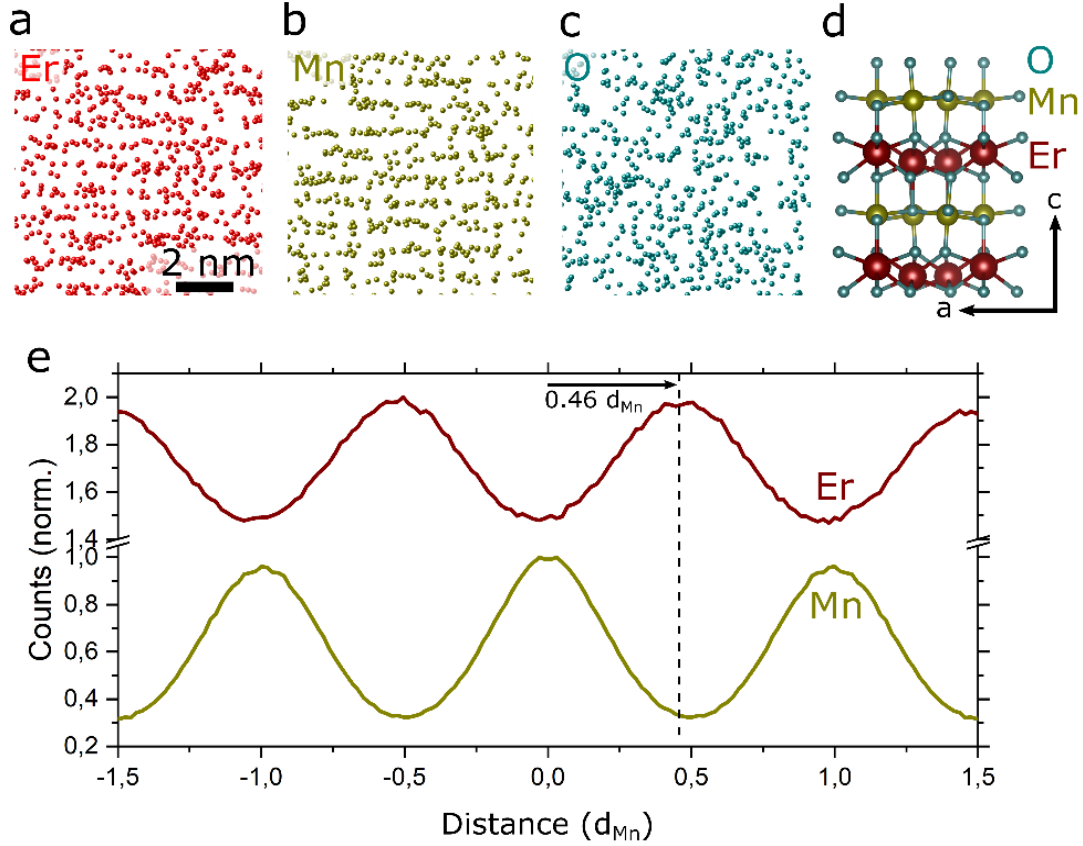

**Figure 4: Atomic resolution in the pole.** (a)-(c) Representation of the atomic planes of the three elements, visualized using the main ionic species (i.e., Er and ErO, Mn, O). (d) ErMnO$_3$ unit cell, illustrating the alternating Er and Mn planes. (e) SDM of the main Er and Mn species (Mn – Mn, Mn – Er and ErO), calculated from a selected area from the pole. Mn is used as reference in all cases, and the horizontal axis is given in terms of the Mn-to-Mn peak distance, d$_{Mn}$. Data is taken at 25K.

This phenomenon will be elaborated in the following using fundamental concepts of field evaporation.

In general, the evaporation rates of Er ($K_{Er}$) and Mn ($K_{Mn}$) need to be equal in order to obtain a stationary state (i.e., $K_{Er} = K_{Mn} = \gamma$ with $\gamma$ a constant) and can be defined by an Arrhenius expression [52,56]:

$$K_i = exp\left(-\frac{Q_i(F_{T,i})}{k_B T}\right) \quad (1)$$

with $i$ the considered atoms (Er or Mn), $k_B$ the Boltzmann constant, $T$ the temperature at the sample surface and $Q_i$ the energy barrier for an atom $i$ at the electric field $F_{T,i}$. This field represents the total electric field above atom $i$ at evaporation and is referred to as the effective evaporation field. The temperature at the sample surface is equal to the analysis temperature $T_0$ (i.e., 25K or 50K, see section 2), in addition to the increase in temperature induced by the laser $\Delta T$. The latter is increasingly important as the laser energy increases. At zero temperature, the value of the electric field at which the energy barrier ($Q_i$) of a specific element is reduced to zero is called the evaporation field ($F_{Ev,i}$). Close to the evaporation field of an atom $i$, the energy barrier is approximated by the expression:

$$Q_i(F_{T,i}) = C_i\left(1 - \frac{F_{T,i}}{F_{Ev,i}}\right) \quad (2)$$

Here, $C_i$ is an energetic constant associated with atom $i$ [57–59]. Experimentally, the linearity of $Q_i(F)$ is approximately observed for some materials in the range of $0.8F_{Ev,i}$ to $0.95F_{Ev,i}$ [60]. Deviations exist outside this regime (i.e., at high laser energy and low $F_{T,i}$), which we neglected in our approach. The energetic constant $C_i$ is a extrapolation of $Q_i(F_i)$ to zero temperature, assuming a linear approximation of $Q_i(F)$, which is valid for fields close to the evaporation field $F_{Ev,i}$. Its value depends on the specific atom (i.e., Mn or Er), but also on the material's composition and structure[13,61,62]. By combining these different expressions, the effective evaporation field ratio of Er ($F_{T,Er}$) and Mn ($F_{T,Mn}$) becomes:

$$\frac{F_{T,Mn}}{F_{T,Er}} = \frac{F_{Ev,Mn}}{F_{Ev,Er}} \times \left(\frac{1 + \frac{k_B T \ln(\gamma)}{C_{Mn}}}{1 + \frac{k_B T \ln(\gamma)}{C_{Er}}}\right) \quad (3)$$

This ratio reflects the amplitude of the local accommodations of the surface curvature during evaporation, which lead to trajectory aberrations and, hence, deviations from the expected inter-atomic distances. At zero temperature, the ratio of the effective fields is equal to that of the evaporation fields. As the temperature increases, the ratio changes depending on the two energetic constants ($C_{Mn}, C_{Er}$). Unfortunately, we do not have values for ErMnO₃ to precisely estimate this ratio (in general, such an estimation is possible based on density functional theory calculations as presented, for example, in Ref. [61–63]). Field evaporation simulations at the atomic scale (see Fig. S1)[64–66], however, indicate the validity of the theoretical expression and demonstrate its influence on the observed offset of Er planes from the correct position.

To visualize the relation between the ratio of the evaporation fields ($F_{T,Mn}/F_{T,Er}$) and the effective one ($F_{Ev,Mn}/F_{Ev,Er}$), and show how it is controlled by the energetic constants ($C_{Er}$ and $C_{Mn}$), we sketch the different scenarios in Figure 6. In Figure 6(a), we show the effective evaporation field as a function of temperature for Mn and Er for three cases: (i) For $C_{Mn} < C_{Er}$, the effective evaporation field of Mn decreases faster than that of Er, leading to an increasing ratio (Figure 6(b)). As a consequence, preferential retention of Er arises, which is substantial at high temperatures, and remains non-zero towards zero Kelvin. This scenario describes our APT analysis of ErMnO₃. (ii) In contrast, for $C_{Mn} > C_{Er}$, the ratio can be reduced to 1 at low temperatures. It grows for higher temperatures, potentially reaching the point where Mn is retained but Er is not. (iii) For $C_{Er} = C_{Mn}$, the ratio of the effective fields is equal to the one of the evaporation fields (at all temperature). It is important to note that this ratio depends on both the element bonding (evaporation field) and temperature and, by extension, on the energy of the laser ($T = T_0 + \Delta T$). The latter implies that the laser energy influences the measured position of the atomic planes (Er offset) as demonstrated in Figure 5, but the preferential retention originates from the distinct chemical bonding of the two elements.

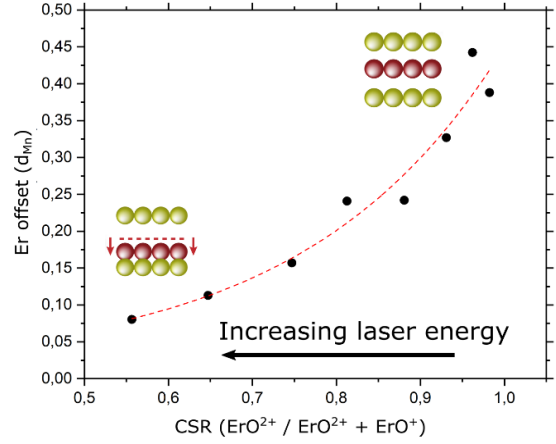

**Figure 5: The Er offset as indicated in Figure 4(e) plotted against the CSR.** The different CSR data points are obtained from the same specimen by varying the laser energy. Atomic offsets are sketched in the plot, where yellow atoms illustrate Mn and red atoms illustrate Er. Data is taken at 50K.

## 4. Conclusion

We investigated the impact of varying operation parameters in APT measurements on the complex oxide ErMnO₃. Consistent with previous studies on other oxide materials, we observed that low-energy laser pulses (i.e., high CSR) result in a higher MRP and a more accurate composition measurement. By analyzing the atomic planes in the (001) pole, we observed an electric field-dependent shift of the Er planes relative to the Mn planes, which does not match the established unit cell structure of $ErMnO_3$. Based on our theoretical analysis and simulations, we attributed this APT-related effect to preferential retention of Er[20].Our results provide new insight into the evaporation behavior of atomic planes in oxide materials with noticeably different evaporation fields and their evolution with temperature, advancing the understanding of atomic-scale APT measurements on complex single phase material systems.

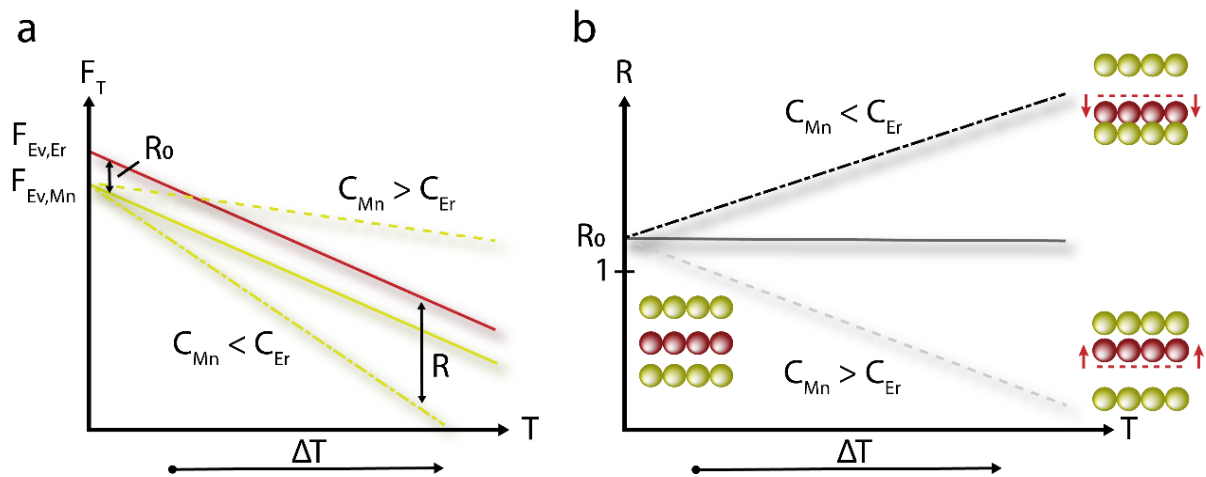

**Figure 6: Temperature dependence of evaporation fields.** (a) Schematic of the change in effective evaporation field, $F_T$, as a function of temperature, T, and the laser induced heating ΔT. The energetic constant, $C_i$, determines the slope and, thus, the ratio, R, between the two elements. (b) Depending on the energetic constant, the ratio between effective evaporation field changes as a function of temperature. At low temperatures, the ratio only depends on the difference in evaporation field, where as at higher temperature the ratio can both increase or decrease. Atomic retention as a result of the change in effective evaporation field ratio is sketched in (b), where atoms (Er – red, Mn – yellow) are displaced from their correct position marked by the dashed line.

## Acknowledgements

The Research Council of Norway (RCN) is acknowledged for the support to the Norwegian Micro- and Nano-Fabrication Facility, NorFab, project number 295864, the Norwegian Laboratory for Mineral and Materials Characterization, MiMaC, project number 269842/F50, and the Norwegian Center for Transmission Electron Microscopy, NORTEM (197405/F50). K.A.H. and D.M. thank the Department of Materials Science and Engineering at NTNU for direct financial support. D.M. acknowledges funding from the European Research Council (ERC) under the European Union's Horizon 2020 research and innovation program (Grant Agreement No. 863691). D.M. thanks NTNU for support through the Onsager Fellowship Program and NTNU Stjerneprogrammet. Hanne-Sofie Søreide is thanked for her support to the APT lab facilities.

## References

[1] P. Dumitraschkewitz, S.S.A. Gerstl, L.T. Stephenson, P.J. Uggowitzer, S. Pogatscher, Clustering in Age-Hardenable Aluminum Alloys, Adv Eng Mater. 20 (2018) 1800255.

[2] B. Gault, A. Chiaramonti, O. Cojocaru-Mirédin, P. Stender, R. Dubosq, C. Freysoldt, S.K. Makineni, T. Li, M. Moody, J.M. Cairney, Atom probe tomography, Nature Reviews Methods Primers. 1 (2021) 1–30.

[3] A.A. El-Zoka, S.H. Kim, S. Deville, R.C. Newman, L.T. Stephenson, B. Gault, Enabling near-atomic–scale analysis of frozen water, Sci Adv. 6 (2020).

[4] A. La Fontaine, A. Zavgorodniy, H. Liu, R. Zheng, M. Swain, J. Cairney, Atomic-scale compositional mapping reveals Mg-rich amorphous calcium phosphate

in human dental enamel, Sci Adv. 2 (2016).

[5] X. Xu, Y. Liu, J. Wang, D. Isheim, V.P. Dravid, C. Phatak, S.M. Haile, Variability and origins of grain boundary electric potential detected by electron holography and atom-probe tomography, Nat Mater. 19 (2020) 887–893.

[6] K.A. Hunnestad, C. Hatzoglou, Z.M. Khalid, P.E. Vullum, Z. Yan, E. Bourret, A.T.J. van Helvoort, S.M. Selbach, D. Meier, Atomic-scale 3D imaging of individual dopant atoms in an oxide semiconductor, Nat Commun. 13 (2022) 4783.

[7] K.A. Hunnestad, J. Schultheiß, A.C. Mathisen, C. Hatzoglou, A.T.J. Van Helvoort, D. Meier, Quantitative 3D mapping of chemical defects at charged grain boundaries in a ferroelectric oxide, 7491 (2022) 1–14.

[8] T. Boll, T. Al-Kassab, Y. Yuan, Z.G. Liu, Investigation of the site occupation of atoms in pure and doped TiAl/$Ti_3$Al intermetallic, Ultramicroscopy. 107 (2007) 796–801.

[9] W.K. Yeoh, B. Gault, X.Y. Cui, C. Zhu, M.P. Moody, L. Li, R.K. Zheng, W.X. Li, X.L. Wang, S.X. Dou, G.L. Sun, C.T. Lin, S.P. Ringer, Direct observation of local potassium variation and its correlation to electronic inhomogeneity in $(Ba_{1-x}K_x)Fe_2As_2$ pnictide, Phys Rev Lett. 106 (2011).

[10] T. Rademacher, T. Al-Kassab, J. Deges, R. Kirchheim, Ordering and site occupancy of D03 ordered $Fe_3Al$ -5 at%Cr evaluated by means of atom probe tomography, Ultramicroscopy. 111 (2011) 719–724.

[11] S. Meher, R. Banerjee, Partitioning and site occupancy of Ta and Mo in Co-base γ/γ' alloys studied by atom probe tomography, Intermetallics (Barking). 49 (2014) 138–142.

[12] C. Cappelli, S. Smart, H. Stowell, A. Pérez-Huerta, Exploring Biases in Atom Probe Tomography Compositional Analysis of Minerals, Geostand Geoanal Res. 45 (2021) 457–476.

[13] C. Hatzoglou, S. Rouland, B. Radiguet, A. Etienne, G. Da Costa, X. Sauvage, P. Pareige, F. Vurpillot, Preferential Evaporation in Atom Probe Tomography: An Analytical Approach, Microscopy and Microanalysis. 26 (2020) 689–698.

[14] D. Zanuttini, I. Blum, L. Rigutti, F. Vurpillot, J. Douady, E. Jacquet, P.M. Anglade, B. Gervais, Simulation of field-induced molecular dissociation in atom-probe tomography: Identification of a neutral emission channel, Phys Rev A (Coll Park). 95 (2017) 061401.

[15] B. Gault, D.W. Saxey, M.W. Ashton, S.B. Sinnott, A.N. Chiaramonti, M.P. Moody, D.K. Schreiber, Behavior of molecules and molecular ions near a field emitter, New J Phys. 18 (2016) 033031.

[16] D.W. Saxey, Correlated ion analysis and the interpretation of atom probe mass spectra, Ultramicroscopy. 111 (2011) 473–479.

[17] M. Müller, B. Gault, G.D.W. Smith, C.R.M. Grovenor, Accuracy of pulsed laser atom probe tomography for compound semiconductor analysis, J Phys Conf Ser. 11 (2011). 012031.

[18] P.A.J. Bagot, O.B.W. Silk, J.O. Douglas, S. Pedrazzini, D.J. Crudden, T.L. Martin, M.C. Hardy, M.P. Moody, R.C. Reed, An Atom Probe Tomography study of site preference and partitioning in a nickel-based superalloy, Acta Mater. 125 (2017) 156–165.

[19] T.J. Prosa, S. Strennen, D. Olson, D. Lawrence, D.J. Larson, A Study of Parameters Affecting Atom Probe Tomography Specimen Survivability,

Microscopy and Microanalysis. 25 (2019) 425–437.

[20] J. Takahashi, K. Kawakami, A quantitative model of preferential evaporation and retention for atom probe tomography, Surface and Interface Analysis. 46 (2014) 535–543.

[21] D.R. Diercks, B.P. Gorman, R. Kirchhofer, N. Sanford, K. Bertness, M. Brubaker, Atom probe tomography evaporation behavior of C-axis GaN nanowires: Crystallographic, stoichiometric, and detection efficiency aspects, J Appl Phys. 114 (2013) 184903.

[22] H. He, J.E. Halpin, S.R. Popuri, L. Daly, J.-W.G. Bos, M.P. Moody, D.A. MacLaren, P.A.J. Bagot, Atom Probe Tomography of a Cu-Doped TiNiSn Thermoelectric Material: Nanoscale Structure and Optimization of Analysis Conditions, Microscopy and Microanalysis. (2021) 1–8.

[23] C. Cappelli, A. Pérez-Huerta, Effect of crystallographic orientation on atom probe tomography geochemical data?, Micron. 137 (2020).

[24] A. Vella, B. Mazumder, G. Da Costa, B. Deconihout, Field evaporation mechanism of bulk oxides under ultra fast laser illumination, J Appl Phys. 110 (2011) 44321.

[25] F. Tang, M.P. Moody, T.L. Martin, P.A.J. Bagot, M.J. Kappers, R.A. Oliver, Practical Issues for Atom Probe Tomography Analysis of III-Nitride Semiconductor Materials, Microscopy and Microanalysis. 21 (2015) 544–556.

[26] F. Li, T. Ohkubo, Y.M. Chen, M. Kodzuka, K. Hono, Quantitative atom probe analyses of rare-earth-doped ceria by femtosecond pulsed laser, Ultramicroscopy. 111 (2011) 589–594.

[27] B. Gault, B. Klaes, F.F. Morgado, C. Freysoldt, Y. Li, F. De Geuser, L.T. Stephenson, F. Vurpillot, Reflections on the Spatial Performance of Atom Probe Tomography in the Analysis of Atomic Neighborhoods, Microscopy and Microanalysis. (2021) 1–11.

[28] T. Prosa, D. Lenz, I. Martin, D. Reinhard, D. Larson, J. Bunton, Evaporation-Field Differences with Deep-UV Atom Probe Tomography, Microscopy and Microanalysis. 27 (2021) 1262–1264.

[29] S.M. Reddy, D.W. Saxey, W.D.A. Rickard, D. Fougerouse, S.D. Montalvo, R. Verberne, A. van Riessen, Atom Probe Tomography: Development and Application to the Geosciences, Geostand Geoanal Res. 44 (2020) 5–50.

[30] E.A. Marquis, B. Gault, Determination of the tip temperature in laser assisted atom-probe tomography using charge state distributions, J Appl Phys. 104 (2008) 84914.

[31] R. Kirchhofer, D.R. Diercks, B.P. Gorman, J.F. Ihlefeld, P.G. Kotula, C.T. Shelton, G.L. Brennecka, Quantifying Compositional Homogeneity in Pb(Zr, Ti)O$_3$ Using Atom Probe Tomography, Journal of the American Ceramic Society. 97 (2014) 2677–2697.

[32] R. Kirchhofer, M.C. Teague, B.P. Gorman, Thermal effects on mass and spatial resolution during laser pulse atom probe tomography of cerium oxide, Journal of Nuclear Materials. 436 (2013) 23–28.

[33] A. Devaraj, R. Colby, W.P. Hess, D.E. Perea, S. Thevuthasan, Role of photoexcitation and field ionization in the measurement of accurate oxide stoichiometry by laser-assisted atom probe tomography, Journal of Physical Chemistry Letters. 4 (2013) 993–998.

[34] D. Santhanagopalan, D.K. Schreiber, D.E. Perea, R.L. Martens, Y. Janssen, P. Khalifah, Y.S. Meng, Effects of laser energy and wavelength on the analysis of LiFePO$_4$ using laser assisted atom

[34] probe tomography, Ultramicroscopy. 148 (2015) 57–66.

[35] D. Zanuttini, I. Blum, L. Rigutti, F. Vurpillot, J. Douady, E. Jacquet, P.M. Anglade, B. Gervais, Simulation of field-induced molecular dissociation in atom-probe tomography: Identification of a neutral emission channel, Phys Rev A (Coll Park). 95 (2017) 061401.

[36] B.B. Van Aken, T.T.M. Palstra, A. Filippetti, N.A. Spaldin, The origin of ferroelectricity in magnetoelectric $YMnO_3$, Nat Mater. 3 (2004) 164–170.

[37] M. Li, H. Tan, W. Duan, Hexagonal rare-earth manganites and ferrites: A review of improper ferroelectricity, magnetoelectric coupling, and unusual domain walls, Physical Chemistry Chemical Physics. 22 (2020) 14415–14432.

[38] S. Artyukhin, K.T. Delaney, N.A. Spaldin, M. Mostovoy, Landau theory of topological defects in multiferroic hexagonal manganites, Nat Mater. 13 (2014) 42–49.

[39] D. Meier, J. Seidel, A. Cano, K. Delaney, Y. Kumagai, M. Mostovoy, N.A. Spaldin, R. Ramesh, M. Fiebig, Anisotropic conductance at improper ferroelectric domain walls, Nat Mater. 11 (2012) 284–288.

[40] J.A. Mundy, J. Schaab, Y. Kumagai, A. Cano, M. Stengel, I.P. Krug, D.M. Gottlob, H. Doğanay, M.E. Holtz, R. Held, Z. Yan, E. Bourret, C.M. Schneider, D.G. Schlom, D.A. Muller, R. Ramesh, N.A. Spaldin, D. Meier, Functional electronic inversion layers at ferroelectric domain walls, Nat Mater. 16 (2017) 622–627.

[41] S.H. Skjærvø, E.T. Wefring, S.K. Nesdal, N.H. Gaukås, G.H. Olsen, J. Glaum, T. Tybell, S.M. Selbach, Interstitial oxygen as a source of p-type conductivity in hexagonal manganites, Nat Commun. 7 (2016).

[42] P. Liu, X.L. Wang, Z.X. Cheng, Y. Du, H. Kimura, Structural, dielectric, antiferromagnetic, and thermal properties of the frustrated hexagonal $Ho_{1-x}Er_xMnO_3$ manganites, Phys Rev B Condens Matter Mater Phys. 83 (2011) 144404.

[43] S.H. Skjærvø, D.R. Småbråten, N.A. Spaldin, T. Tybell, S.M. Selbach, Oxygen vacancies in the bulk and at neutral domain walls in hexagonal $YMnO_3$, Phys Rev B. 98 (2018) 184102.

[44] Z. Yan, D. Meier, J. Schaab, R. Ramesh, E. Samulon, E. Bourret, Growth of high-quality hexagonal $ErMnO_3$ single crystals by the pressurized floating-zone method, J Cryst Growth. 409 (2015) 75–79.

[45] K. Thompson, D. Lawrence, D.J. Larson, J.D. Olson, T.F. Kelly, B. Gorman, In situ site-specific specimen preparation for atom probe tomography, Ultramicroscopy. 107 (2007) 131–139.

[46] C. Hatzoglou, Norwegian Atom Probe App, (2023).

[47] T. Choi, Y. Horibe, H.T. Yi, Y.J. Choi, W. Wu, S.W. Cheong, Insulating interlocked ferroelectric and structural antiphase domain walls in multiferroic $YMnO_3$, Nat Mater. 9 (2010) 253–258.

[48] M.E. Holtz, K. Shapovalov, J.A. Mundy, C.S. Chang, Z. Yan, E. Bourret, D.A. Muller, D. Meier, A. Cano, Topological Defects in Hexagonal Manganites: Inner Structure and Emergent Electrostatics, Nano Lett. 17 (2017) 5883–5890.

[49] J.A. Mundy, J. Schaab, Y. Kumagai, A. Cano, M. Stengel, I.P. Krug, D.M. Gottlob, H. Doğanay, M.E. Holtz, R. Held, Z. Yan, E. Bourret, C.M. Schneider, D.G. Schlom, D.A. Muller, R. Ramesh, N.A. Spaldin, D. Meier, Functional electronic inversion layers at ferroelectric domain walls, Nat Mater. 16 (2017) 622–627.

[50] D.R. Kingham, The post-ionization of field evaporated ions: A theoretical explanation of multiple charge states, Surf Sci. 116 (1982) 273–301.

[51] N. Amirifar, R. Lardé, E. Talbot, P. Pareige, L. Rigutti, L. Mancini, J. Houard, C. Castro, V. Sallet, E. Zehani, S. Hassani, C. Sartel, A. Ziani, X. Portier, Quantitative analysis of doped/undoped ZnO nanomaterials using laser assisted atom probe tomography: Influence of the analysis parameters, J Appl Phys. 118 (2015) 215703.

[52] B. Gault, M. Moody, J. Cairney, S. Ringer, Atom probe tomography, Springer, New York, NY, 2012.

[53] B. Gault, F. Danoix, K. Hoummada, D. Mangelinck, H. Leitner, Impact of directional walk on atom probe microanalysis, Ultramicroscopy. 113 (2012) 182–191.

[54] B.P. Geiser, T.F. Kelly, D.J. Larson, J. Schneir, J.P. Roberts, Spatial distribution maps for atom probe tomography, Microscopy and Microanalysis. 13 (2007) 437–447.

[55] F. Vurpillot, A. Bostel, E. Cadel, D. Blavette, The spatial resolution of 3D atom probe in the investigation of single-phase materials, Ultramicroscopy. 84 (2000) 213–224.

[56] E.W. Müller, Field desorption, Physical Review. 102 (1956) 618–624.

[57] M.K. Miller, R.G. Forbes, Atom probe tomography, Elsevier, 2009.

[58] Williams Lefebvre, F. Vurpillot, X. Sauvage, Atom Probe Tomography, Elsevier, London, 2016.

[59] R.G. Forbes, Field evaporation theory: a review of basic ideas, Appl Surf Sci. 87–88 (1995) 1–11.

[60] G.L. Kellogg, Measurement of activation energies for field evaporation of tungsten ions as a function of electric field, Phys Rev B. 29 (1984) 4304–4312.

[61] T. Carrasco, J. Peralta, C. Loyola, S.R. Broderick, Modeling field evaporation degradation of metallic surfaces by first principles calculations: A case study for Al, Au, Ag, and Pd, J Phys Conf Ser. 1043 (2018) 012039.

[62] J. Peralta, S.R. Broderick, K. Rajan, Mapping energetics of atom probe evaporation events through first principles calculations, Ultramicroscopy. 132 (2013) 143–151.

[63] T. Ohnuma, Surface Diffusion of Fe and Cu on Fe (001) under Electric Field Using First-Principles Calculations, Microscopy and Microanalysis. 25 (2019) 547–553.

[64] F. Vurpillot, A. Cerezo, D. Blavette, D.J. Larson, Modeling image distortions in 3DAP, Microscopy and Microanalysis. 10 (2004) 384–390.

[65] M. Gruber, F. Vurpillot, A. Bostel, B. Deconihout, Field evaporation: A kinetic Monte Carlo approach on the influence of temperature, Surf Sci. 605 (2011) 2025–2031.

[66] D. Blavette, F. Vurpillot, P. Pareige, A. Menand, A model accounting for spatial overlaps in 3D atom-probe microscopy, Ultramicroscopy. 89 (2001) 145–153.

[67] C. Hatzoglou, S. Rouland, B. Radiguet, A. Etienne, G. Da Costa, X. Sauvage, P. Pareige, F. Vurpillot, Preferential Evaporation in Atom Probe Tomography: An Analytical Approach, Microscopy and Microanalysis. 26 (2020) 689–698.

# Supplementary information for
# Optimizing compositional and atomic-level information of oxides in atom probe tomography

**Authors:** K. A. Hunnestad[1], C. Hatzoglou[1], F. Vurpillot[5], I-E. Nylund[1], Z. Yan[2,3], E. Bourret[3], A. T. J. van Helvoort[4], D. Meier[1*]

[1]Department of Materials Science and Engineering, Norwegian University of Science and Technology (NTNU), 7491 Trondheim, Norway

[2]Department of Physics, ETH Zurich, Zürich, Switzerland

[3]Materials Sciences Division, Lawrence Berkeley National Laboratory, Berkeley, CA, USA

[4]Department of Physics, Norwegian University of Science and Technology (NTNU), 7491 Trondheim, Norway

[5]UNIROUEN, CNRS, Groupe de Physique des Matériaux, Normandie Université, Av. de l'Université BP12, 76801 St Etienne du Rouvray, France

*Corresponding author. Email: dennis.meier@ntnu.no

The simulation model used in this study has been developed at the Groupe de Physique des Materiaux (GPM) by Vurpillot et al. More details of this model and its applications can be found in [64–67]. To reproduce the studied system as faithfully as possible, we have simulated the field evaporation of a successive stack of atomic planes of Er and Mn, using a simplified cubic structure. To account for the thermally active process of the evaporation, a Monte-Carlo type algorithm chooses the next atom to evaporate according to its evaporation rate (equations 1 and 2). However, for this simulation model, the energy constant C, used to calculate the energy barrier height (expression 2) is the same independent of the chemical nature of the atoms (i.e., $C_{Er} = C_{Mn}$). To date, there is no evaporation model that takes this energy constant into account, therefore we have studied the influence of the evaporation field ratio ($F_{Ev,Mn}/F_{Ev,Er}$ from 0.65 to 1.35), as well as the temperature (50 and 250K) on the measured offset between atomic Er and Mn planes (same protocol as used experimentally i.e., SDM), as shown in Figure S1.

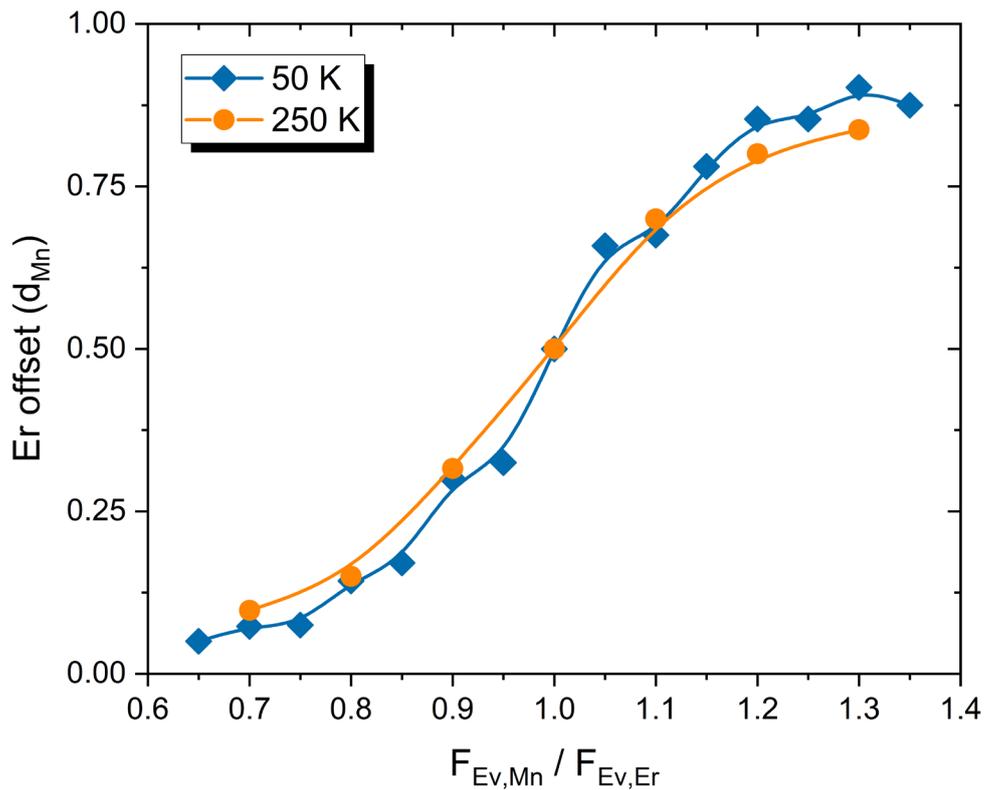

**Figure S1**: Er Offset ($d_{Mn}$) as a function of the simulated evaporation field ratio $F_{Ev,Mn}/F_{Ev,Er}$ for two temperatures 50 and 250 K.

First, there is no clear influence of the temperature on the measured Er offset, as expected form equation 3, when Er and Mn have the same energetic constant C (i.e., $C_{Er} = C_{Mn}$). Second, there is a strong influence of the evaporation field ratio on the measured offset. Depending on whether the atomic planes of Er are low field ($F_{Ev,Mn}/F_{Ev,Er} > 1$) or high field ($F_{Ev,Mn}/F_{Ev,Er} < 1$) compared to Mn atomic planes, the deviation to the expected offset ($0.5d_{Mn}$) is respectively either positive or negative. A positive deviation ($> 0.5d_{Mn}$) means that the reconstructed Er atomic planes are closer to previously evaporated Mn atomic planes rather than to the Mn planes that are next in line to be evaporated.. For the negative deviation, ($< 0.5 d_{Mn}$), we can observe the opposite occurring. In addition, the deviation (positive or negative) increases as the field ratios increase in magnitude. These simulation results clearly indicate that the evaporation field ratios, and thus the effective field ratio (since there is no influence of the temperature and all the energetic constants are equal), generate a deviation from the expected value of the offset.